\documentclass[aps,prb,superscriptaddress,citeautoscript,floatfix,twocolumn]{revtex4-2}
\usepackage{amsmath, amsthm}
\usepackage{amsfonts}
\usepackage{graphicx}
\usepackage{dcolumn}
\usepackage{bm}
\usepackage{lipsum}
\usepackage{physics}
\usepackage[caption=false]{subfig}
\usepackage{graphicx}
\usepackage{ulem}

\usepackage[colorlinks=true,bookmarks=false,citecolor=blue,linkcolor=red,hyperfootnotes=true,urlcolor=blue]{hyperref}

\newcommand{\veck}{\mathbf{k}}

\begin{document}
\title{Physically interpretable approximations of many-body spectral functions}
\author{Shubhang Goswami}
\email{sgoswam3@illinois.edu}
\affiliation{Department of Physics, University of Illinois Urbana-Champaign, Urbana, Illinois 61801, USA}

\author{Kipton Barros}
\affiliation{Theoretical Division and CNLS, Los Alamos National Laboratory, Los Alamos, New Mexico 87545, USA}

\author{Matthew R. Carbone}
\affiliation{Computational Science Initiative, Brookhaven National Laboratory, Upton, New York 11973, USA}
\date{\today}

\begin{abstract}
The rational function approximation provides a natural and interpretable representation of response functions such as the many-body spectral functions. We apply the Vector Fitting (VFIT) algorithm to fit a variety of spectral functions calculated from the Holstein model of electron-phonon interactions. We show that the resulting rational functions are highly efficient in their fitting of sharp features in the spectral functions, and could provide a means to infer physically relevant information from a spectral dataset. The position of the peaks in the approximated spectral function are determined by the location of poles in the complex plane. In addition, we developed a variant of VFIT that incorporates regularization to improve the quality of fits. With this new procedure, we demonstrate it is possible to achieve accurate spectral function fits that vary smoothly as a function of physical conditions.

\end{abstract}

\maketitle

\section{Introduction}
The spectral function $A(\mathbf{k}, \omega)$ provides rich information about a many-body system, and is directly measurable in angle-resolved photoemission spectroscopy (ARPES)~\cite{damascelli2003angle}. Theoretically, the spectral function is obtained from the many-body Green's function i.e., $A(\veck, \omega) = -\frac{1}{\pi} \Im G(\veck, \omega).$ Due to causality, the spectral function can also be used to reconstruct the Green's function via Kramers-Kronig relations, providing a bidirectional relationship between theory and experiment.

To obtain the many-body spectral function, one generally calculates the many-body Green's function. There are myriad ways to approximate the many-body Green's function, as solving for the exact Green's function is equivalent to diagonalizing the fully interacting Hamiltonian, the complexity of which scales exponentially in the number of electrons. Some of these approximations include dynamical mean-field theory (DMFT)~\cite{kotliarDMFT2006}, the GW approximation~\cite{hedin1965GWA}, the momentum average (MA) family of methods \cite{berciu2006Greens,goodvin2006Greens,Berciu_2010} including the generalized Green’s function cluster expansion (GGCE)~\cite{carbone2021numerically,carbone2021bond,carbone2022generalized}, continuous time quantum Monte-Carlo (CT-QMC)~\cite{gullCTMC2011,rubtsov2005continuous,cohen2014green}, and Green's function Monte-Carlo (GFMC)~\cite{kalos1962GFMC,schmidt2005GFMC,ceperley1984quantum}. Even machine learning and data-driven methods have been employed to directly predict and analyze spectral functions~\cite{arsenaultAIMML2014,sturmspectralML2021,milesKondo2021,andrews2023selfsimilarity}.

Interpreting spectral functions poses system-specific challenges that require both experimental methods and theoretical insights. To address this, our approach provides low-dimensional interpretable features obtained from data that capture the essential information of spectral functions. Specifically, we parameterize spectral data using a rational function approximation, i.e., as a {\it ratio} of two polynomials. Our chosen ansatz is in the form of a simple-pole expansion, $A(\omega) \approx \sum_j \varphi_j/(z-\xi_j),$ where the complex-valued fitting parameters $\xi_j$ and $\varphi_j$ correspond to the poles and residues respectively. If the residues $\varphi_j$ were purely real, then the approximation would become a sum of Lorentenzian functions, with positions and widths determined by the real and imaginary parts of $\xi_j,$ respectively. By employing a rational function representation we can concisely capture singular features commonly found in real world data and naturally interpret them via the fitting parameters.

Rational functions can be made to conform to the analytic structure of spectral functions. For example, because $A(\omega)$ is real, all poles $\xi_j$ appearing in its simple pole expansion must appear as complex-conjugate pairs. By retaining only the poles in the lower half plane, it is straightforward to analytically continue from $A(\omega)$ to the retarded Green's function for the entire upper-half of the complex plane, which is appropriately non-singular. Also, rational approximation can readily achieve an exact integral sum rule (via a constraint on the discrete sum over residues), and the correct $1/\omega$ asymptotic decay of $A(\omega)$ at large frequencies. New rational approximation schemes may therefore be of interest for the extremely challenging problem of numerical analytic continuation~\cite{schott2016analytic,ferris-prabhu_numerical_1973,motoyama2022robust,Han2017analytic,Gull2021analytic,Gull2021nevanlinna,Gull2023analytic} or for other tasks such as the numerical renormalization group~\cite{ziga2013pade}, and calculation of self-energies~\cite{farid_luttinger-ward_2021,farid1992semiconductors,faridGW1991}. Finally, better representations of spectral data may be a starting point for the development of machine learning models, e.g., that interpolate spectral function data over a range of physical regimes. Whereas many previous studies focused on predicting $A(\omega)$ for each value of $\omega$ separately~\cite{sturmspectralML2021,milesKondo2021,arsenaultAIMML2014}, future models could aim to predict the few poles and residues needed to encode the entire spectral function, thus leading to large efficiency gains. The resulting models are readily suitable for input to many-body calculations such as obtaining the density of states or running self-consistent calculations.

In this work, we demonstrate that the Vector Fitting (VFIT)~\cite{gustavsen1999rational,gustavsen1998simulation} algorithm can efficiently and robustly perform these fits. Furthermore, through an appropriate regularization scheme, we show that poles and residues for a {\it collection} of fits can be made to vary relatively smoothly as a function of model system physical parameters.

\section{Methods \label{sec:methods}}

Rational function approximation assumes the fitting form,
\begin{equation}
    r(z) = p(z) / q(z),\label{eq:rational}
\end{equation}
i.e., a ratio of polynomials $p(z)$ and $q(z)$. Because $r(z)$ is intended to fit a spectral function $A(\omega)$, and the latter decays like $|\omega|^{-1}$ for $\omega \rightarrow \pm \infty$, the polynomial degree of $p(z)$ should be one less than that of $q(z)$. Then, assuming the roots of  $q(z)$ non-degenerate, $r(z)$ can be expressed as a simple pole expansion,
\begin{equation}
r(z) = \sum_{j=1}^m \varphi_j/(z-\xi_j).\label{eq:simple pole}
\end{equation}
One can verify that Eq.~\eqref{eq:rational} is recovered with $q(x)= \prod_j (z-\xi_j)$ and $p(z) = r(z) q(z)$, which are indeed polynomials of the correct order.  When approximating a real function on the real axis, $z=\omega$, the terms in the simple pole expansion (both poles $\xi_j$ and residues $\varphi_j$) should come in complex conjugate pairs.

The presence of poles $\xi_j$ near the real axis result in sharp features in $r(z)$, that may be controlled through careful tuning of the fitting parameters $\varphi_j$ and $\xi_j$. This property makes rational approximation particularly well-suited for approximating functions with singular, or near-singular, behaviors. In the field of computational physics, this approach finds applications in various numerical scenarios. For example, it can be used to approximate the discontinuous step in a zero-temperature Fermi function~\cite{sidje2011rational,moussa2016minimax} and to handle polynomial divergences encountered in lattice gauge theory~\cite{clark2006rational}.

This paper is focused on the approximation of spectral functions $r(\omega) \approx A(\omega)$ for real-valued frequencies $\omega$. Spectral functions frequently contain sharp, Lorentzian-like peaks that correspond to quasi-particle excitations. The rational function ansatz, Eq.~\eqref{eq:rational}, is a perfect fit to this application, because a single pole $\xi_j$ (and its complex conjugate) can exactly model a Lorentzian. The width of the peak is tunable by varying the distance of the poles from the real axis, $\pm \mathrm{Im}~\xi_j$, while the weight of the peak is tunable through the real part of the residue, $\mathrm{Re}~\varphi_j.$
Also, the simple pole expansion $r(z)$ makes it easy to perform analytic continuation. For a given approximation $r(\omega) \approx A(\omega)$, one can get the same result by defining a new simple pole expansion $\tilde{G}(z)$ that discards poles in the upper half of the complex plane, and multiplies the remaining residues by $-2\pi i$. By construction, $-\frac{1}{\pi} \Im \tilde{G}(\omega) = r(\omega)$, and  $\tilde{G}(z)$ has the analytic structure expected of the retarded Green function.

To find a rational approximation $r(z)$ to a given function $f(z)$, sampled at a given set of points $z \in \{\lambda_i\}_{i=1}^{N}$, we may seek to minimize the squared error,
\begin{equation}
    J = \sum_{i=1}^N \left[f(\lambda_i) - r(\lambda_i)\right]^2.\label{eq:J def}
\end{equation}
The VFIT~\cite{gustavsen1999rational,gustavsen1998simulation} algorithm iteratively refines a rational approximation $r(z)$ to minimize Eq.~\eqref{eq:J def}. A key idea is to employ the barycentric fitting form,
\begin{equation}
    r(z) = \frac{\sum _{j=1}^m \varphi_j / (z - \xi_j)}{1 + \sum _{j=1}^m \psi_j / (z - \xi_j)}.\label{eq:barycentric}
\end{equation}
Relative to Eq.~\eqref{eq:simple pole}, there are additional fitting parameters $\psi_j$. These are, in some sense, redundant. Indeed, one can return to Eq.~\eqref{eq:rational} by multiplying both the numerator and denominator by $\prod_{j=1}^m (z - \xi_j)$. Nonetheless, the presence of $\psi_j$ will be expedient to the fitting procedure. Convergence of VFIT is characterized by the condition $\psi_j \rightarrow 0$, and this limit coincides with the simple pole expansion of Eq.~\eqref{eq:simple pole}.

The strategy of VFIT is to alternate between two types of updates steps: Step 1 updates the residues $\varphi_j$ and $\psi_j$ to minimize the fitting error. Step 2 updates the interpolation nodal points $\xi_j$ to match the poles of the currently-fitted rational approximation.

We will start by describing Step 2. The update rule selects $\xi_j \rightarrow \xi'_j$ as the zeros of $1 + \sum_{j=1}^m \psi_j / (z - \xi_j)$. The motivation is to modify the $\xi_j$ such that the subsequent refitting of $\varphi_j$ and $\psi_j$ will (hopefully) bring all $\psi_j$ closer to zero. To build intuition for why this might happen, suppose the current $r(z)$ of Eq.~\ref{eq:barycentric} is already a good rational approximation. Hypothetically, if we were to update $\xi_j \rightarrow \xi'_j$ and simultaneously set $\psi_j \rightarrow 0$, then we would obtain a new rational approximation in the form of a simple pole expansion, in which the poles from $r(z)$ are unchanged. In practice, however, coefficients $\varphi_j$ and $\psi_j$ will be updated by refitting to the data. For more discussion on the $\xi_j$ updates, see Ref.~\onlinecite{gustavsen1999rational}.

Step 1 of VFIT fits the coefficients $\varphi_j$ and $\psi_j$ to optimize agreement $f(\lambda_i) \approx r(\lambda_i)$ for each data point $i$ using a convenient error measure. Using the barycentric fitting form, the desired agreement may be written
\begin{equation} \label{eq: VFIT linear coefficient equation}
    f(\lambda_i) \left(1 + \sum_{j=1}^m \frac{\psi_j}{\lambda_i-\xi_j} \right) \approx \sum_{j=1}^m\frac{\varphi_j}{\lambda_i-\xi_j}.
\end{equation}
In matrix notation, this becomes $\mathbf{b} \approx A \mathbf{x}$, where
\begin{align}
    A &=    \begin{bmatrix}
           \frac{1}{\lambda_1 - \xi_1} & \cdots & \frac{1}{\lambda_1 - \xi_m} & \frac{-f(\lambda_1)}{\lambda_1 - \xi_1} & \cdots & \frac{-f(\lambda_1)}{\lambda_1 - \xi_m}\\
           \vdots &&\vdots & \vdots && \vdots\\
           \frac{1}{\lambda_N - \xi_1} & \cdots & \frac{1}{\lambda_N - \xi_m} &
           \frac{-f(\lambda_N)}{\lambda_N - \xi_1} & \cdots & \frac{-f(\lambda_N)}{\lambda_N - \xi_m}
        \end{bmatrix} \\
    \mathbf{x} &= [\varphi_1, \dots , \varphi_m, \psi_1, \dots , \psi_m]^T \\
    \mathbf{b} &= [f(\lambda_1), \dots , f(\lambda_N)]^T. \\
\end{align}

In the original VFIT, the unknown parameters $\mathbf{x}$ were solved in a least-squares sense. In our work, we define an intermediate objective function that includes an $L_2$ regularization term,
\begin{equation}
    \tilde{J} = \abs{\mathbf{b} - A\mathbf{x}}^2 + \gamma \abs{\mathbf{x}}^2.
\end{equation}
The minimizer is
\begin{equation}
    \mathbf{x}=(A^TA + \gamma I)^{-1} A^T\mathbf{b},
    \label{eq: VFIT matrix equation with regularizer}
\end{equation}
where $I$ is the $2m\times 2m$ identity. The regularization strength $\gamma$ will typically be small, and penalizes large values of $\varphi_j$ and $\psi_j$. The elements of $\mathbf{x}$ are used to update $\varphi_j, \psi_j$. This completes Step 1 of VFIT.

These two steps of VFIT should be applied iteratively until the fitting parameters $\varphi_j, \psi_j, \xi_j$ converge. Convergence of the pole update implies $\psi_j \rightarrow 0$, and the barycentric approximation $r(z)$ takes the form of a simple pole expansion, Eq.~\eqref{eq:simple pole}, that minimizes the squared error, Eq.~\eqref{eq:J def}. Note that VFIT is not guaranteed to converge, and if it converges, it may reach only a local minimum of $\tilde{J}.$ The convergence properties of VFIT are still a subject of debate~\cite{lefteriu2013convergence,shi2016nonconvergence}, and some convergence rate analysis is provided in Ref.~\onlinecite{drmac_vector_2015}. Other rational approximation algorithms exist that may address some of the limitations of VFIT~\cite{berljafa_rkfit_2017,nakatsukasa2018aaa,xu2013bootstrap,gugercin2008h_2}. In the present study, we will demonstrate that we can usually get good fits by annealing the regularization strength $\gamma$ through three values: $10^{-3} \rightarrow 10^{-6} \rightarrow 0$, and employing 33 VFIT iterations for each $\gamma$ value.

\section{Results}

\begin{figure}
    \centering
    \includegraphics[width=0.48\textwidth]{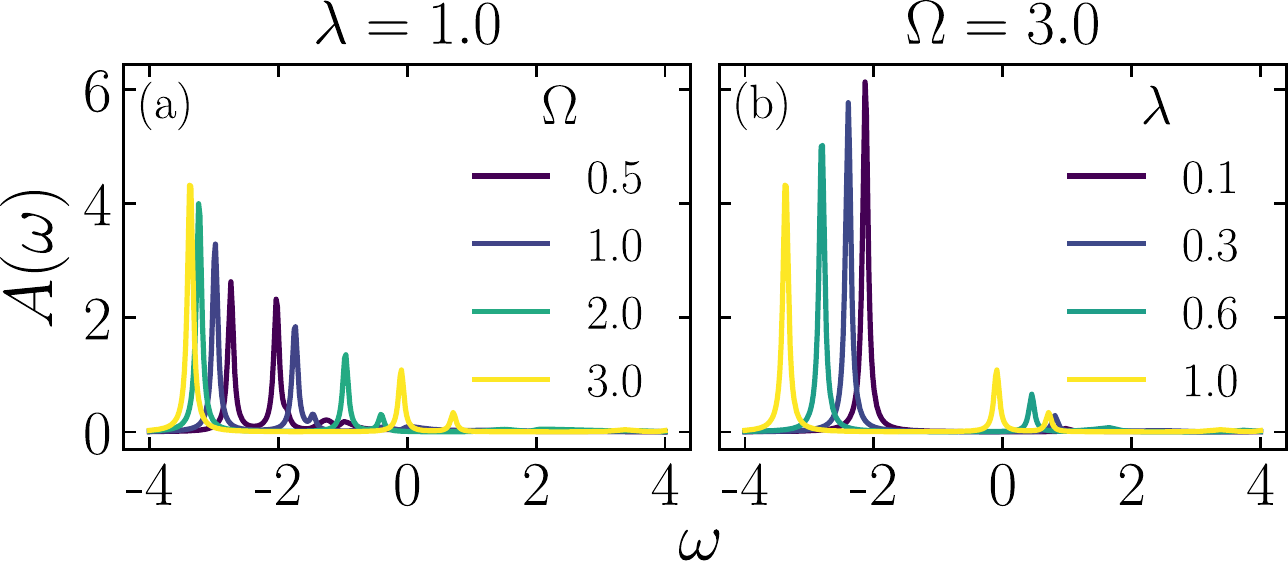}
    \caption{A sample of Holstein spectral functions $A(\omega)$ with (a) varying phonon frequencies $\Omega$ or (b) varying coupling strengths $\lambda$.}
    \label{fig:BNLData colorplot}
\end{figure}

To demonstrate our method, we fit a collection of spectral data of the Holstein model simulated using the exact Green's function cluster expansion method developed by Carbone~\textit{et al.}~\cite{carbone2021numerically}. The Holstein model is given by the Hamiltonian
\begin{equation}
    \hat{H} = -t\sum_{\langle ij \rangle} \hat{c}_i^\dagger \hat{c}_j + \Omega \sum_i \hat{b}_i^\dagger \hat{b}_i + \hat{V}_\mathrm{H}  
\end{equation}
where, $t$ is the fermion hopping coefficient, $\Omega$ is the phonon frequency and $\hat{V}_\mathrm{H}$ is the fermion-boson interaction potential of the Holstein model, given by,
\begin{equation}
    \hat{V}_\mathrm{H} = \alpha \sum_i \hat{c}_i^\dagger \hat{c}_i(\hat{b}_i^\dagger + \hat{b}_i).
\end{equation}
A dimensionless coupling parameter is given by $\lambda$ which is the ratio of the ground-state energy in the atomic limit to that of the free particle limit, i.e.,
\begin{equation}
    \lambda = \frac{E_{0}(t=0)}{E_{0}(\alpha=0)} = \frac{2\alpha^2}{\Omega t}.
\end{equation}

To showcase the rational approximation, we fit a dataset of spectral functions with varying dimensionless coupling parameter $\lambda$ and phonon frequency $\Omega.$ We sampled 100 values of each parameter in equal spaced intervals over the ranges $0.1 \leq \lambda \leq 1$ and $0.5 \leq \Omega \leq 3$, yielding a total of $10^4$ spectral functions. Each spectral function was calculated at 400 evenly spaced intervals at frequency values $\omega\in [-4, 4].$
A sample of these spectra is shown in Fig.~\ref{fig:BNLData colorplot}. 

\begin{figure}
    \centering
    \includegraphics[width=0.49\textwidth]{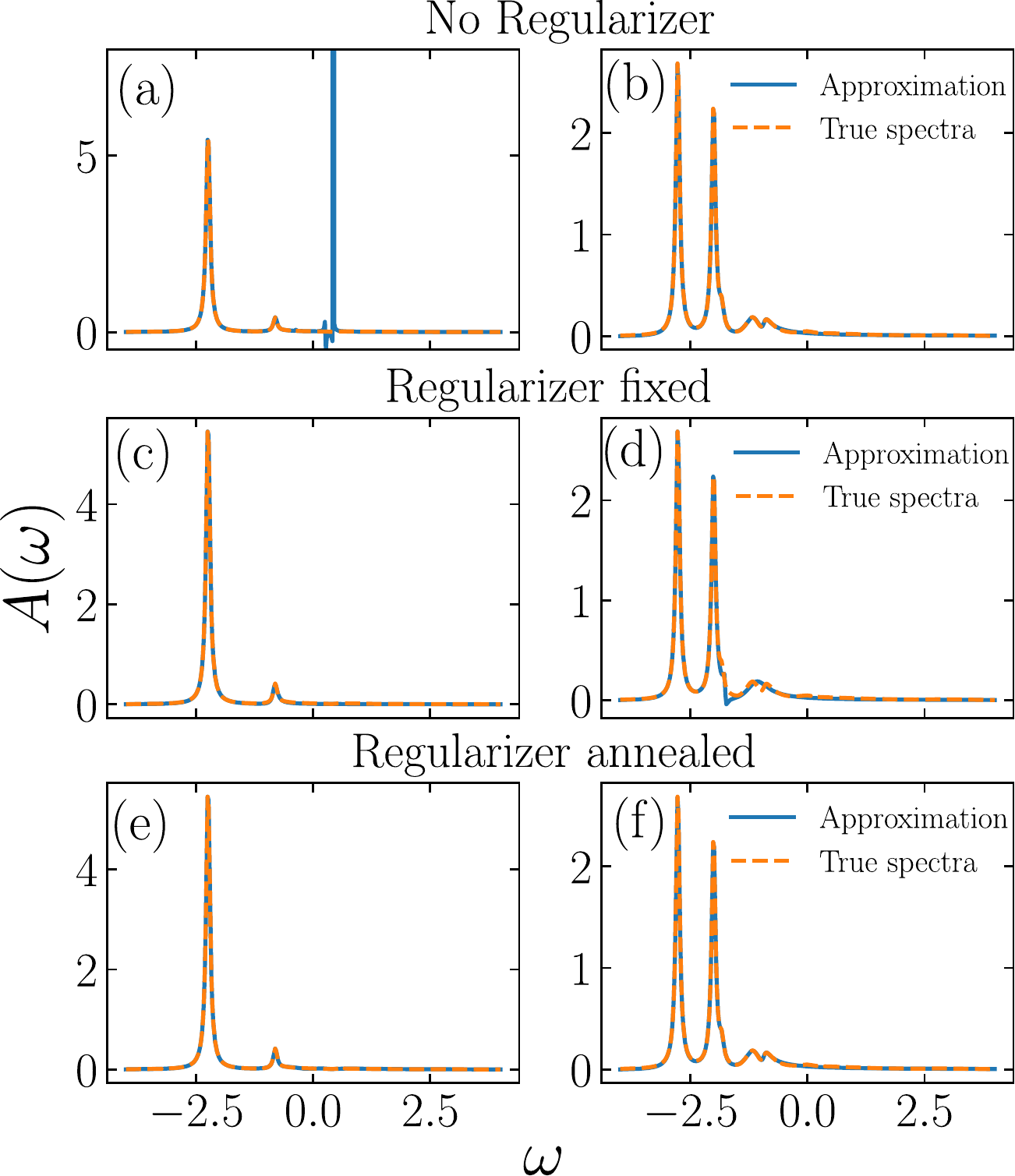}
    \caption{To improve the quality of spectral function fits, we incorporated regularization into the VFIT algorithm, as described in the main text. Left and right columns show two different fitting tasks. Each spectral function was fit using $5$ poles, and their complex conjugates. Top row: the original VFIT algorithm may converge to a bad local minimum of the objective function. Middle row: VFIT with a fixed regularization strength of $\gamma = 10^{-3}$ converges more reliably, but the fit is sub-optimal. Bottom row: Annealing the regularization strength $\gamma$ to zero yields the best fits.}
    \label{fig:annealing_justification}
\end{figure}

Figure~\ref{fig:annealing_justification} demonstrates the importance of including an annealed regularization procedure into VFIT. For illustrative purposes, we selected two spectral functions with different numbers of peaks. In all cases, we used the fitting form of Eq.~\ref{eq:barycentric} with $m=10$ that corresponds to 5 poles and their complex conjugates, which allows for 5 peaks. There is significantly more flexibility than needed to obtain good fits of the reference curve appearing in Fig.~\ref{fig:annealing_justification} (a,c,e), as discussed in Appendix A.  The original VFIT algorithm (no regularization, $\gamma=0$) is compared to VFIT with a fixed regularization strength ($\gamma=10^{-3}$), and to VFIT where $\gamma$ is annealed in three stages ($10^{-3} \rightarrow 10^{-6} \rightarrow 0$) over $100$ VFIT iterations. Each spectral function was fit independently, starting from randomized initialization of the fitting parameters $\varphi_j, \psi_j, \xi_j.$ Our protocol for annealing the regularization strength is effective in escaping local minima and finding globally good fits. Specifically, the procedure tends to avoid spurious spikes in the fitted spectral functions when there are more fitting parameters than necessary. We will continue use this annealed regularization protocol in the remainder of this paper.

\begin{figure}
    \centering
    \includegraphics[width=0.49\textwidth]{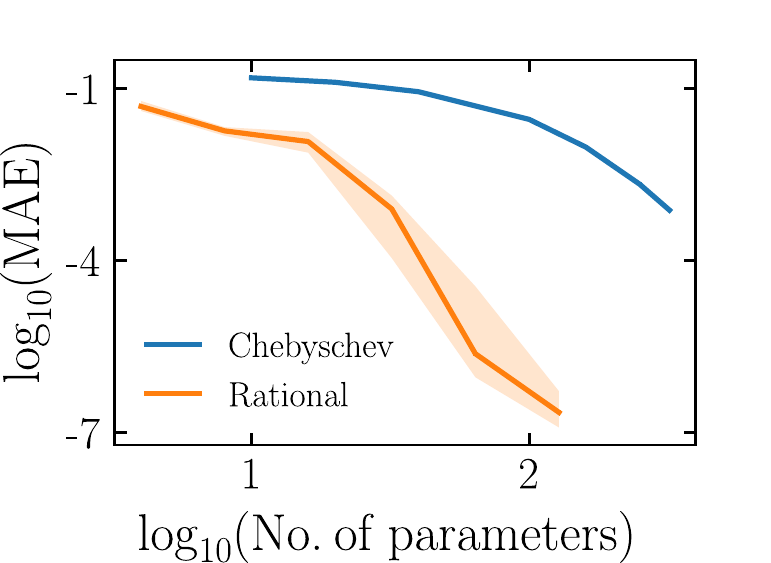}
    \caption{The mean absolute error (MAE) of spectral function fits at coupling strength $\lambda=1$ and averaged over phonon frequencies $\Omega=0.5$ to $3$. Each spectral function was fit using 5 poles and their complex conjugates. The $x$-axis counts the number of independent real parameters; for rational fits, these are the real and imaginary parts of the poles and residues in Eq.~\eqref{eq:simple pole}, not counting complex conjugate pairs. The line denotes the median of the MAE for spectral functions from $\Omega=0.5$ to $3$ for a fixed $\lambda=1$ and the shaded confidence bands indicate the region between the 25th and 75th percentile. For this dataset of spectral functions, rational approximation needs fewer parameters than Chebyschev polynomial expansion.}
    \label{fig:VFIT vs Chebyschev errors}
\end{figure}

In the past, spectral functions have been approximated using Chebyshev polynomial fitting~\cite{hendry2021chebyschev,vanasse62fejer}. We demonstrate that rational approximation using regularized VFIT is a far better choice for approximating the sharp features of spectral functions. We applied VFIT to approximate 100 spectral functions of the Holstein model with phonon frequency $\Omega \in [0.5, 3]$ at $\lambda=1.$ As shown in Fig.~\ref{fig:VFIT vs Chebyschev errors}, fitting spectral functions with rational functions requires far fewer parameters than using Chebyshev polynomial approximation to achieve a similar degree of accuracy, as measured by mean absolute error (MAE). 

\begin{figure}
    \centering
    \includegraphics[width=0.49\textwidth]{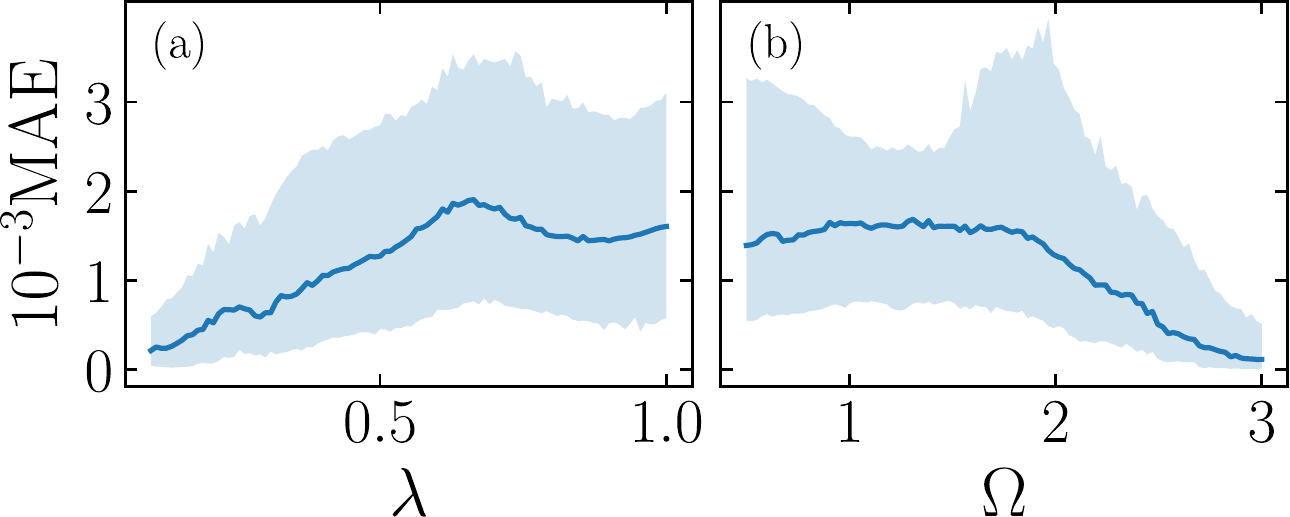}
    \caption{Each spectral function was fit using $5$ poles and their complex conjugates (a) The mean absolute error of rational approximations for each value of $\lambda$ averaged over $\Omega=0.5$ to $3$. (b) The mean absolute error of rational approximations for each value of $\Omega$ averaged over $\lambda=0.1$ to $1$. The shaded regions in both (a) and (b) represent the range of the mean absolute error between the 25th and 75th percentile. These results are largely independent of the initial guess for the fitting parameters.}
    \label{fig:VFIT_MAE}
\end{figure}

The mean absolute errors over the entire dataset are described by the two panels in Fig.~\ref{fig:VFIT_MAE}. Figure~\ref{fig:VFIT_MAE}a shows the MAE of 100 spectra at each coupling $\lambda$ whereas Fig.~\ref{fig:VFIT_MAE}b shows the MAE of 100 spectra at each value of the phonon frequency $\Omega$. Note that in general, it is more challenging to converge spectra generated by the exact Green's function cluster expansion method corresponding to larger values of $\lambda$ and smaller values of $\Omega$~\cite{carbone2021numerically}.

The rational approximation method also provides a heuristic for determining the number of parameters for the rational function. For instance, spectral functions of the Holstein model at high $\lambda$ and low $\Omega$ have 5 unique peaks. A reasonable choice is to employ 5 poles in the lower-half of the complex plane, as well as their complex conjugates to capture each of the respective peaks. Recall that the spectral function we are fitting is purely real, so both the poles and residues must come in complex conjugate pairs. 

\begin{figure}
    \centering
    \includegraphics[width=0.49\textwidth]{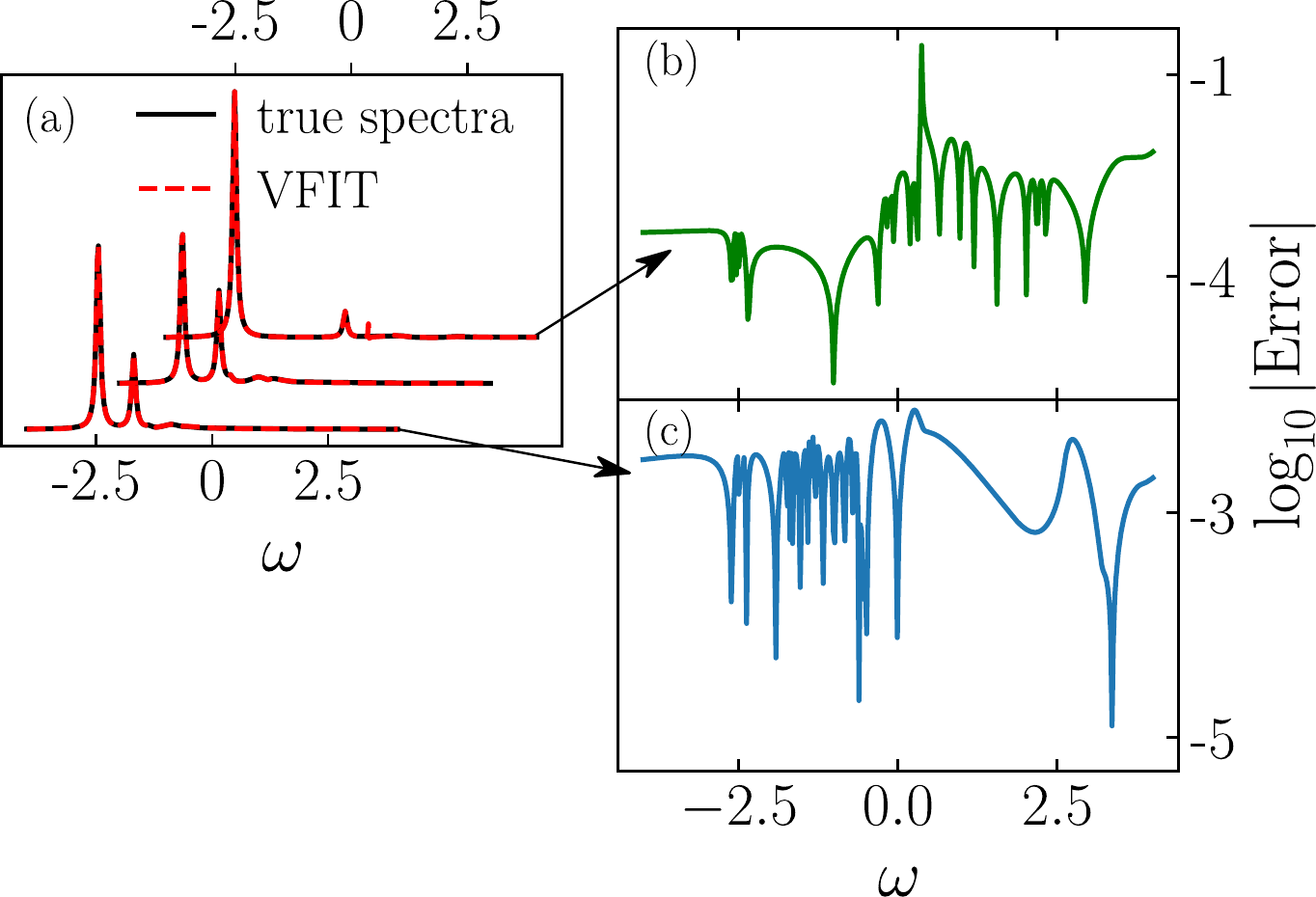}
    \caption{(a) Typical spectral function fits. Each spectral function was fit using $5$ poles and their complex conjugates. The three spectral functions were selected according to their MAE. From top to bottom, these are the fits with errors in the 75th, 50th, and 25th percentile. Each spectral function and its fit was shifted horizontally and vertically for improved visibility. The bottom x axis corresponds to the fit of the 25th percentile and the top x-axis corresponds to the fit of the 75th percentile (b, c) Errors for the 75th and 25th percentile fits, respectively.}
    \label{fig:VFIT_waterfall}
\end{figure}

In Figure~\ref{fig:VFIT_waterfall}, the approximations representing the first, second, and third MAE quartiles are presented. A sharp, unphysical spike is visible in the third fit (75th percentile error), and is associated with large imaginary values of the residues of the rational approximation. Such fitting errors could typically be eliminated by using an even slower annealing of the regularization strength $\gamma$. For the present study, however, we will continue to use only $10^2$ regularized VFIT iterations per spectral function.

\begin{figure}
    \centering
    \includegraphics[width=0.49\textwidth]{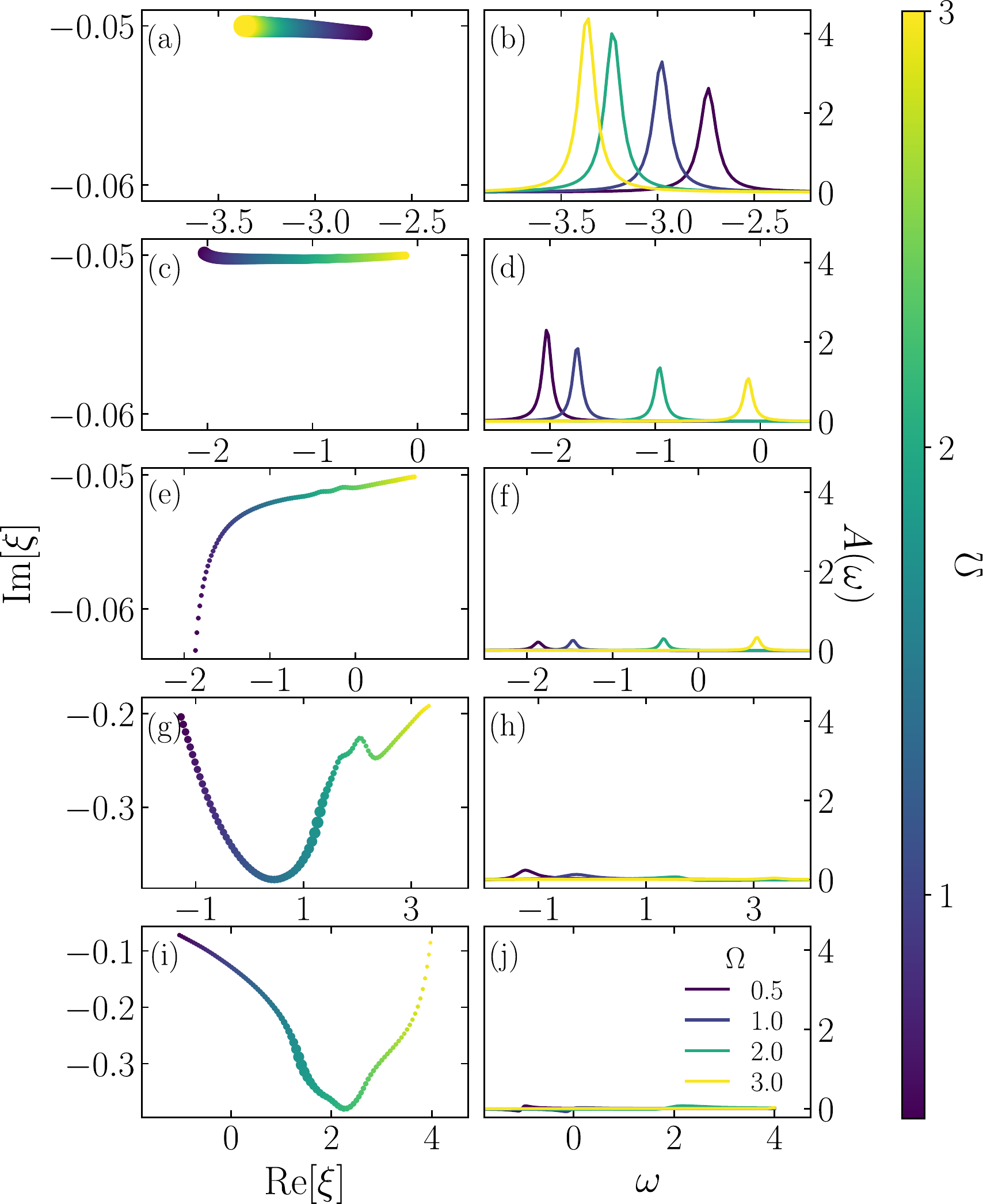}
    \caption{Visualization of the fitted simple pole expansion for spectral functions with $\lambda=1.0$ and varying $\Omega$. Left: Positions of the 5 poles in the negative half of the complex plane. The real and imaginary axes control peak location and width, respectively. Marker size indicates the magnitude of the corresponding residues, which control spectral weight and the shape of each peak. Right: The reconstructed contribution of each pole to the spectral fit; for purely real residues, these contributions would be exactly Lorentzian. }
    \label{fig:VFIT smooth poles omega colorbar contribution}
\end{figure}

The parameters learned by VFIT frequently admit physical interpretation. The location of each peak corresponds to a quasi-particle excitation energy, the width of the peak to its lifetime, and the total area under the curve to its weight. In Fig.~\ref{fig:VFIT smooth poles omega colorbar contribution}, the poles and their associated contributions to the rational approximation are displayed at different $\Omega$ values. From the figure it is evident that the poles achieved by VFIT are responsible for each of the quasi-particle excitations seen in the spectral function and they vary smoothly with changes in $\Omega$. 

Furthermore, the poles achieved by the rational approximation algorithm are continuous functions of the parameters that generated them. This means that as the parameters are varied, the poles vary continuously as well, giving a direct connection between the quasi-particle energies and the physical parameters of the model. This is shown in Fig.~\ref{fig:VFIT smooth poles omega colorbar contribution} for varying phonon frequency $\Omega$. At weaker couplings $\lambda \lesssim 0.5,$ the spectral functions lack sufficient structure to warrant the use of 5 poles and their complex conjugates. As shown in Appendix~\ref{sec:smaller}, two unique poles can be effective for this case.

\section{Conclusion}
We have demonstrated that rational function approximation is a good choice for fitting spectral functions. Using a modified version of the VFIT algorithm, we find that such fits can be produced efficiently and fairly reliably. In our variant of VFIT, we have incorporated a Tikhonov regularization scheme. The strength of this regularization can be gradually reduced during the fitting process. The rational approximation comes in the form of a simple pole expansion. The resulting poles and residues are physically interpretable as quasi-particle energies and lifetimes. These outputs can potentially be used as feature sets for future machine learning-based physics applications. The poles and residues learned by VFIT capture the essential information of the spectral function, which can be used to reconstruct and interpolate the spectral function. This is useful because it allows us to perform further analysis on the spectral function, such as computing various physical properties. The poles and residues can also be used to calculate other quantities such as the density of states. Additionally, the fact that the poles and residues are continuous functions of the parameters that generated them makes it possible to study how the spectral function changes as these parameters are varied. 

The power of rational approximation goes beyond just parameterizing spectral functions. The advantage of using rational functions to approximate sharply changing functions can be applied in a variety of domains, such as modeling circuit responses~\cite{chou2021equivalent,swaminathan2010designing} or simulating spike trains in neuroscience~\cite{palmieri2015transfer,stewart2009spiking}. In the context of self-consistent calculations, Von Barth and Holm demonstrated the accuracy of representing spectral functions as a sum of Gaussians~\cite{holm1998fully}. However, using a sum of Lorentzians could be a more natural choice. By restricting the residues to be real, VFIT can be used to automatically decompose spectral functions as a sum of Lorentzians. It would be interesting to compare the accuracy of using rational approximation in self-consistent loops with the traditional sum of Lorentzians or Gaussians. In conclusion, rational approximation using VFIT not only provides an effective tool for parameterizing and analyzing spectral function but also holds potential for broader applications in various scientific domains.

\section{Code Availability}
The code used for generating the rational approximation fits in this paper along with a sample data file are provided here \href{https://github.com/ShubhangG/Rational-Approximation-for-many-body-spectral-functions}{github.com/ShubhangG/Rational-Approximation-for-many-body-spectral-functions}.

We also produced a small package in Julia which users can use to generate approximations for any functions, which is available at \href{https://github.com/ShubhangG/VFitApproximation}{github.com/ShubhangG/VFitApproximation}.

\begin{acknowledgements}
S.G acknowledges Los Alamos National Laboratory for supporting his graduate research internship. This research is based upon work supported by the U.S. Department of Energy, Office of Science, Office Basic Energy Sciences, under Award Numbers DE-SC0022311, FWP PS-030 and DE-SC0012704.
\end{acknowledgements}

\appendix

\section{smaller set of poles} \label{sec:smaller}

\begin{figure}
    \centering
    \includegraphics[width=0.49\textwidth]{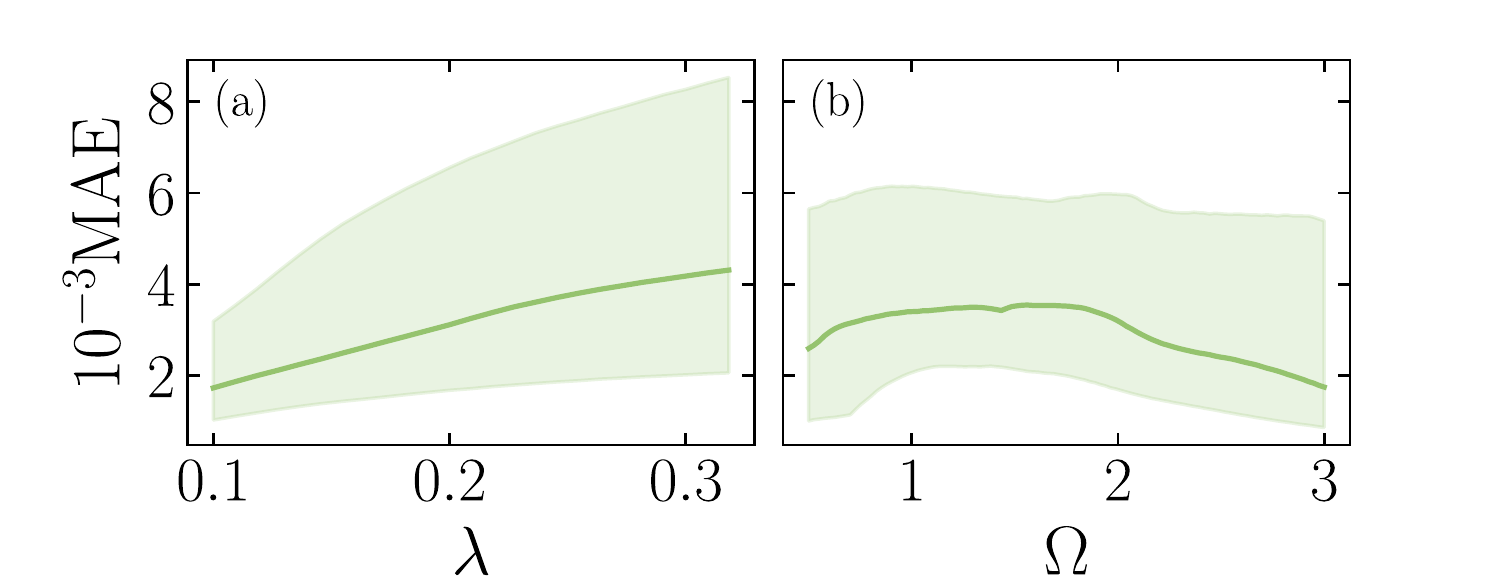}
    \caption{(a) The mean absolute error of rational approximations with 2 poles and their complex conjugates for each value of $\lambda$ averaged over $\Omega=0.5$ to $3$. (b) The mean absolute error of rational approximations for each value of $\Omega$ averaged over $\lambda=0.1$ to $0.33$. The shaded regions in both (a) and (b) represent the range of the mean absolute error between the 25th and 75th percentile }
    \label{fig:MAE_2poles_small_coupling}
\end{figure}

\begin{figure}
    \centering
    \includegraphics[width=0.49\textwidth]{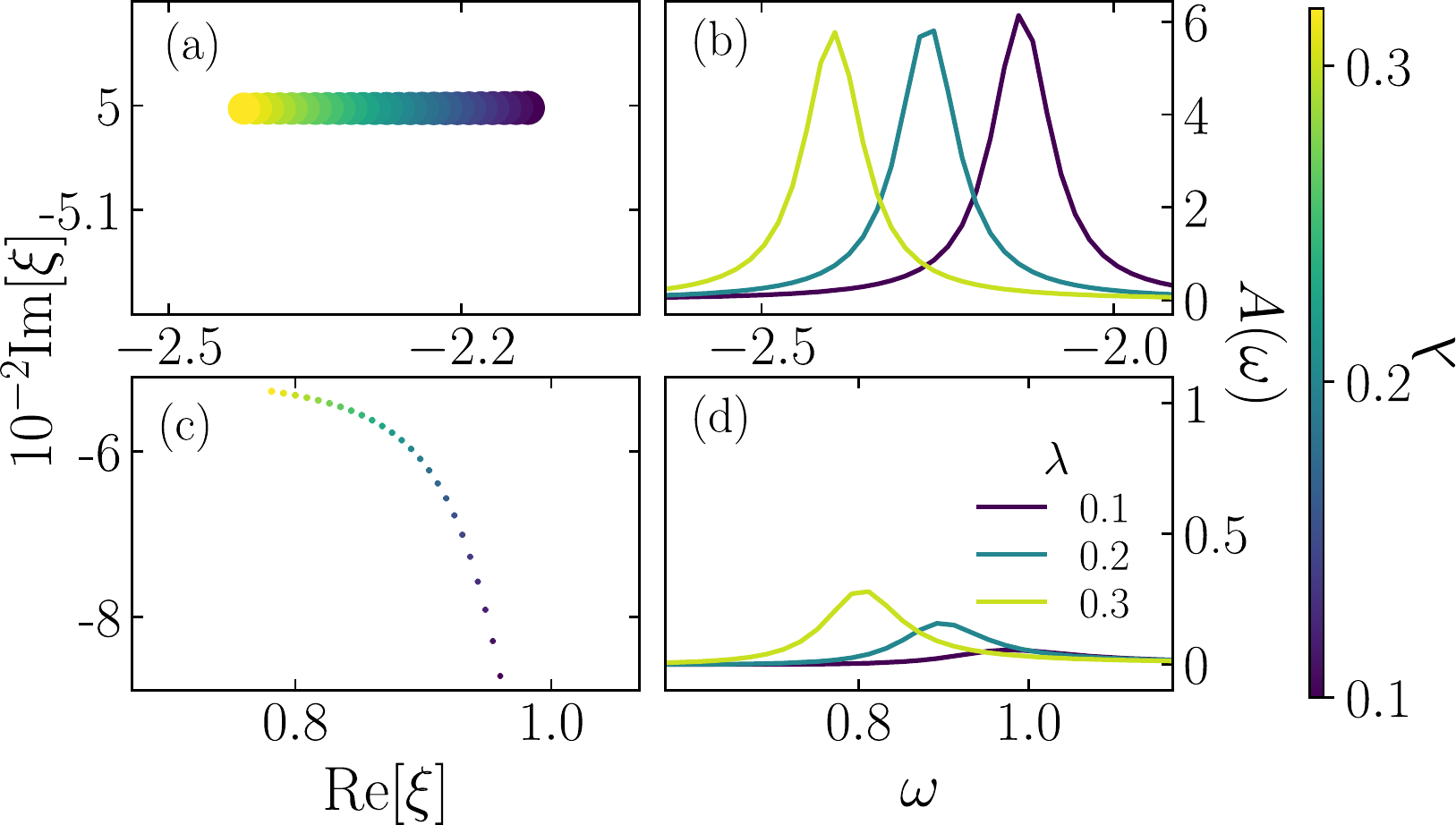}
    \caption{Visualization of the fitted simple pole expansion for spectral functions with $\Omega=3.0$ and varying $\lambda$. Left and right columns should be interpreted analogously to those in Fig.~\ref{fig:VFIT smooth poles omega colorbar contribution}.}
    \label{fig:small_coupling_2_poles}
\end{figure}

We have empirically observed that a good spectral function fit can be achieved by selecting the number of independent poles equal to the number of peaks. Spectral functions of the Holstein model with low coupling $\lambda \lesssim 0.5$, have only 2 peaks. Correspondingly, we expect to achieve a good fit with only 2 unique poles and their complex conjugates. 
This is verified in Fig.~\ref{fig:MAE_2poles_small_coupling}, which shows the quartiles of the MAE for these fits at $\Omega=3.0$ and $0.1 < \lambda < 0.33$.
Here, we disabled regularization (i.e., set $\gamma = 0$) as the original VFIT algorithm demonstrated reliable convergence.

Figure~\ref{fig:small_coupling_2_poles} shows the contributions of the two poles to these fits. The panels are analogous to those in Fig.~\ref{fig:VFIT smooth poles omega colorbar contribution}, but here we required fewer poles to get a good fit.

\section{Effect of noise}
\begin{figure}
    \centering
    \includegraphics[width=0.49\textwidth]{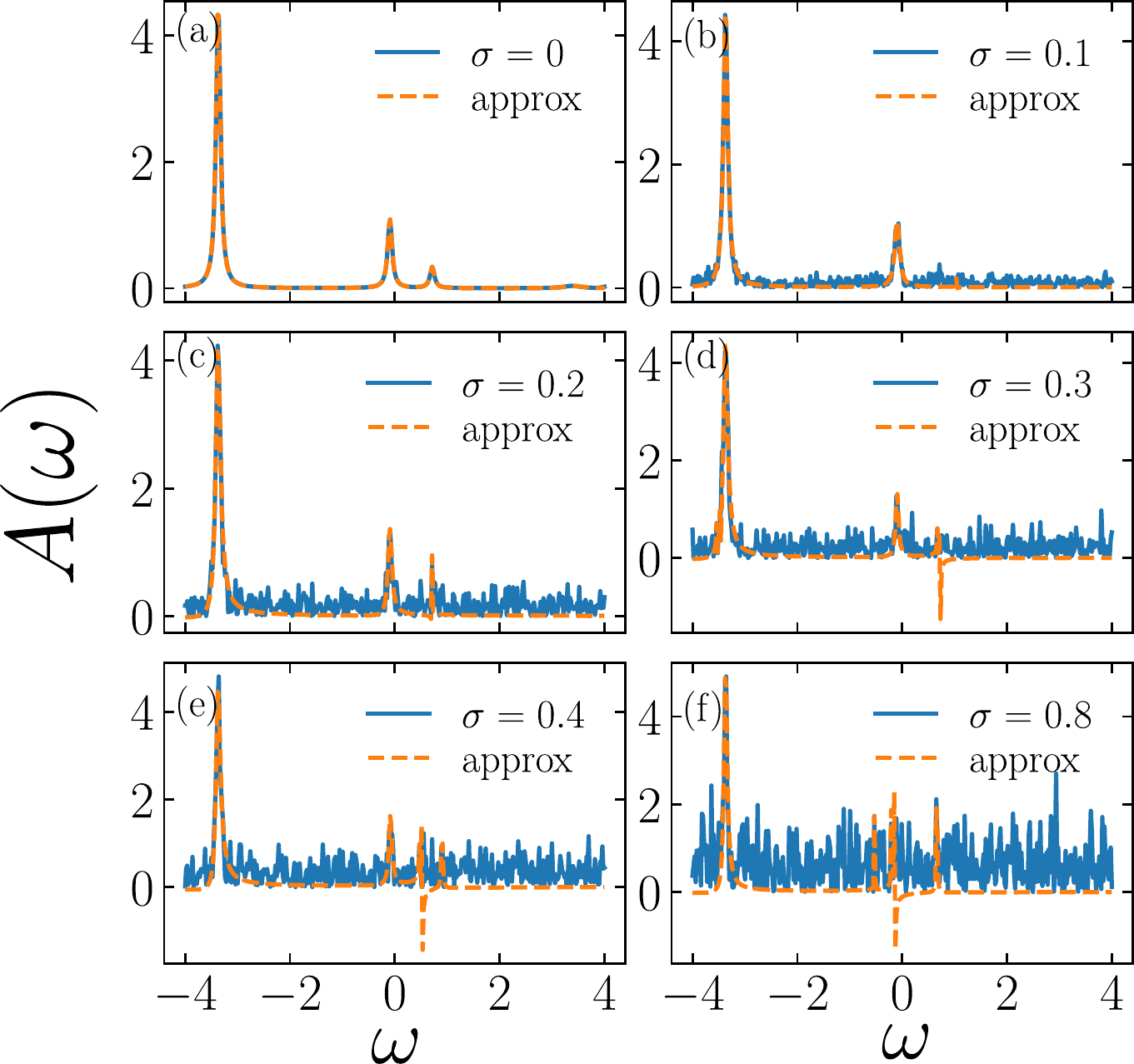}
    \caption{Fitting a spectral function under increasing noise. Each spectral function was fit using 4 poles and their complex conjugates. Gaussian square noise was used with mean 0 and increasing variance as seen in (a)-(f) given by $\sigma$.}
    \label{fig:noise-analysis}
\end{figure}

We conducted a small assessment of how our method does in presence of noise. In fig~(\ref{fig:noise-analysis}), we took a simple spectra with distinct peaks of different sizes and added gaussain square noise with mean 0 but increasing variance given by the sub-figure legends. We tried fitting the spectra with 4 poles and their complex conjugates. It is known that for noisy measurements, the traditional VFIT algorithm iteration may exhibit no convergence or may have multiple points of convergence~\cite{lefteriu2013convergence}

We observed that in cases where noise becomes prominent and obscures subtle peaks, the poles that were previously fitting the now obscured peak tend to start fitting the sharp noise. However, even under these conditions, the algorithm remains effective in capturing distinct peaks whose signal strength surpasses the noise. For instance, the prominent peak near $\omega=-4$ remains discernible, even in the presence of exceedingly high noise. The second most significant peak vanishes only when the noise overwhelms the signal. In such situations, our heuristic of choosing the number of poles based on the count of visible peaks will be fruitful. As long as the peaks remain distinguishable, we have confidence that our algorithm can successfully detect them, although we cannot guarantee convergence.

\bibliographystyle{apsrev4-1}
\bibliography{references}

\end{document}